\documentclass{article}
\usepackage{portland}
\usepackage{lscape}
\usepackage{amsfonts}
\usepackage{amsmath}
\usepackage{geometry}

\input{tcilatex}
\begin{document}

\title{Integration of Third Order Ordinary Differential Equations Possessing
Two-Parameter Symmetry Group by Lie's Method }
\author{Mladen Nikoli\'{c} and Milan Rajkovi\'{c}\thanks{%
E-mail: milanr@vin.bg.ac.yu} \\
Institute of Nuclear Sciences Vin\v{c}a,\\
11001 Belgrade, Serbia}
\maketitle

\begin{abstract}
The solution of a class of third order ordinary differential equations
possessing two parameter Lie symmetry group is obtained by group theoretic
means. It is shown that reduction to quadratures is possible according to
two scenarios: 1) if upon first reduction of order the obtained second order
ordinary differential equation besides the inherited point symmetry acquires
at least one more new point symmetry (possibly a hidden symmetry of Type
II). 2) First, reduction paths of the fourth order differential equations
with four parameter symmetry group leading to the first order equation
possessing one known (inherited) symmetry are constructed. Then, reduction
paths along which a third order equation possessing two-parameter symmetry
group appears are singled out and followed until a first order equation
possessing one known (inherited) symmetry are obtained. The method uses
conditions for preservation, disappearance and reappearance of point
symmetries.
\end{abstract}


\section{Introduction}

Providing a unified treatment the Lie group theory has developed into a
powerful tool for solving and classifying differential equations even though
the theory does not apply to all equations. In order to broaden and
complement the already existing results various group theoretic approaches
have been devised relying, for example, on hidden\cite{Barbara1}A, \cite%
{Barbara2}, and nonlocal symmetries\cite{Nonlocal}, \cite{Geronimi}. In this
paper we use hidden and convertible symmetries in order to expand the
results related to solutions of third order differential equations
possessing two parameter symmetry group. We define a convertible type of
symmetry of order $n-1$ as a point symmetry that disappears during the first
reduction of order of an ordinary differential equation, remains hidden
(non-local) during $n-1$ reductions, and reappears as a point symmetry after 
$n$ reductions. Convertible symmetries may be regarded as a special class of
hidden symmetries of type II\cite{Barbara1}. Integration of third order
differential equations which admit a three-dimensional solvable and
non-solvable symmetry algebra has been previously discussed\cite{Ibragimov}.
ln general, third order ordinary differential equations possessing two
parameter symmetry group are not solvable. However, due to properties of
hidden or convertible symmetries in certain cases this may be possible.
Since two parameter group is always solvable one symmetry generator may be
used to reduce the order of the initial equation so that the other symmetry
is inherited as a local symmetry of the corresponding second order equation.
This property has been effectively used in order to obtain the solution of a
class of second-order ordinary differential equations not possessing Lie
point symmetries\cite{GovinderLeach}. In this approach, the order of the
initial second order equation is increased and the solution of the new third
order equation is sought in the case that additional symmetries (for example
hidden symmetries of type II) appear both in the new third order equation as
well as in the reduced second order equation.

For third order differential equations possessing two point symmetries there
are two possibilities for the complete reduction i.e. reduction to
quadratures:

\textit{Case 1}. If upon first reduction of order, the obtained second order
ordinary differential equation besides the inherited point symmetry
possesses at least one more new point symmetry which could be a type II
hidden symmetry.

\textit{Case 2}. Following the first reduction of order the inherited point
symmetry is the only point symmetry of the obtained second order
differential equation. Further reduction using that symmetry generates a
first order equation with one known (inherited) point symmetry so that the
initial third order equation may be solved (first order ordinary
differential equations have infinite number of point symmetries however
generally unknown). The idea is to use properties of hidden (convertible)
symmetries that come up along the reduction paths of the fourth order
differential equations with four parameter symmetry group that lead to the
first order equation possessing one known symmetry. Conditions for the
appearance and disappearance of hidden symmetries used in the analysis
pertaining to this Case as well as the role of three-dimensional subalgebra
in the analysis of hidden symmetries are presented in our recent paper\cite%
{Nikolic}. Naturally, this method does not enable the complete
classification of third order equations possessing two parameter symmetry
group since the source of hidden symmetries may be in equations of order
higher than fourth, as well as in contact symmetries.

Although some aspects of Case 1 have been previously discussed in relation
to the second order equations not possessing Lie point symmetries\cite%
{GovinderLeach}, we give a detailed account of it for completeness. Hence, a
separate section of the paper is devoted to each case, and the reduction
paths pertaining to the four-parameter symmetry groups are presented in the
Appendix A.

\section{Case 1: The reduced second order equation possesses an additional,
second symmetry}

The starting point of the analysis is a third order differential equation%
\begin{equation}
y^{\prime \prime \prime }=F(x,y,y^{\prime },y^{\prime \prime }),  \label{0}
\end{equation}%
possessing two parameter Lie point symmetry group, which is reduced to a
second order equation using one of the available symmetries. According to
the Lie's classification there are four two-dimensional transitive algebras
of vector fields in $\mathbb{R}^{2}$\cite{Lie}:\renewcommand{%
\arraystretch}{1.3}%
\begin{equation*}
\begin{tabular}{|l|l|l|l|}
\hline
Type & $L_{2}$ structure & \multicolumn{2}{|l|}{Basis of $L_{2}$ in
canonical variables} \\ \hline
I & $[X,Y]=0$ & $X=\frac{\partial }{\partial y}$ & $Y=\frac{\partial }{%
\partial x}$ \\ \hline
II & $[X,Y]=0$ & $X=\frac{\partial }{\partial y}$ & $Y=x\frac{\partial }{%
\partial y}$ \\ \hline
III & $[X,Y]=X$ & $X=\frac{\partial }{\partial y}$ & $Y=y\frac{\partial }{%
\partial y}$ \\ \hline
IV & $[X,Y]=X$ & $X=\frac{\partial }{\partial y}$ & $Y=x\frac{\partial }{%
\partial x}+y\frac{\partial }{\partial y}$ \\ \hline
\end{tabular}%
\end{equation*}%
\begin{equation*}
\text{{\small Table 1. Canonical form of two-dimensional Lie algebras.} }
\end{equation*}%
In the following exposition we consider each possibility.

\subsection{Type I 
\protect\begin{tabular}{l}
$X=\partial _{y},\text{ \ }Y=\partial _{x}$%
\protect\end{tabular}%
}

The general form of a differential equation invariant under the action of
symmetries%
\begin{equation}
X=\partial _{y},\text{ \ \ \ \ }Y=\partial _{x}.  \label{1}
\end{equation}%
is%
\begin{equation}
y^{\prime \prime \prime }=F(y^{\prime },y^{\prime \prime }).  \label{2}
\end{equation}%
If eq. (\ref{2}) is reduced using vector field $X$ (setting $u=x$ and $%
v=y\prime $), a second order differential equation is obtained%
\begin{equation}
v^{\prime \prime }=F(v,v^{\prime }),  \label{3}
\end{equation}%
or in terms of the canonical coordinates%
\begin{equation}
u^{\prime \prime }=\bar{F}(v,u^{\prime })=-u^{\prime 3}F(v,\frac{1}{%
u^{\prime }}).  \label{4}
\end{equation}%
Equation (\ref{3}) has an inherited symmetry%
\begin{equation*}
\tilde{Y}^{(1)}=\partial _{u},
\end{equation*}%
so it may be further reduced to the first order differential equation. In
our notation tilde denotes restriction of the symmetry generator to
corresponding local coordinates (differential invariants), while superscript
index in brackets denotes the order of the prolongation. However in order to
completely reduce eq. (\ref{2}) to quadratures, we suppose that equation (%
\ref{3}) possesses an additional symmetry $\tilde{Z}$ which could be a
hidden symmetry of type II. Now, there are also four possibilities for $%
\tilde{Y}^{(1)}$ and $\tilde{Z}:$\renewcommand{\arraystretch}{1.3}

\begin{equation*}
\begin{tabular}{|c|c|c|c|c|}
\hline
\multicolumn{1}{|c|}{Type} & \multicolumn{1}{|c|}{$[\tilde{Y}^{(1)},$ $%
\tilde{Z}]$} & \multicolumn{2}{|l|}{\ \ \ \ \ \ \ \ Vector field} & 
Differential equation \\ \hline
A & $0$ & $\tilde{Y}^{(1)}=\partial _{u}$ & $\tilde{Z}=\partial _{v}$ & $%
u^{\prime \prime }=\bar{F}(u^{\prime })$ \\ \hline
B & $0$ & $\tilde{Y}^{(1)}=\partial _{u}$ & $\tilde{Z}=v\partial _{v}$ & $%
u^{\prime \prime }=\bar{F}(v)$ \\ \hline
C & $\tilde{Y}^{(1)}$ & $\tilde{Y}^{(1)}=\partial _{u}$ & $\tilde{Z}%
=u\partial _{u}+v\partial _{v}$ & $vu^{\prime \prime }=\bar{F}(u^{\prime })$
\\ \hline
D & $\tilde{Z}$ & $\tilde{Y}^{(1)}=\partial _{u}$ & $\tilde{Z}=u\partial
_{u} $ & $u^{\prime \prime }=u^{\prime }\bar{F}(v)$ \\ \hline
\end{tabular}%
\end{equation*}%
\begin{equation*}
\text{ {\small Table 2.} {\small Canonical forms of}\ }{\small u}^{{\small %
\prime \prime }}{\small =\tilde{F}(v,u}^{{\small \prime }}{\small )}\text{ 
{\small invariant under }}{\small \tilde{Y}}^{{\small (1)}}\text{{\small and}
}{\small \tilde{Z}.}
\end{equation*}

\subsubsection{Type IA}

Comparison of eq. (\ref{4}) with the canonical form of equation
corresponding to Type A shows that this case would imply the existence of
three point symmetries for eq. (\ref{2}) so that it does not belong to the
class considered here.

\subsubsection{Type IB}

For this type the canonical form of differential equation is%
\begin{equation*}
u^{\prime \prime }=\bar{F}(v),
\end{equation*}%
so that%
\begin{equation*}
-u^{\prime 3}F(v,\frac{1}{u^{\prime }})=\bar{F}(v),
\end{equation*}%
or%
\begin{equation*}
F(v,\frac{1}{u^{\prime }})=-\frac{1}{u^{\prime 3}}\bar{F}(v).
\end{equation*}%
Recalling that%
\begin{equation*}
u^{\prime }=\frac{1}{v^{\prime }},
\end{equation*}%
yields 
\begin{equation*}
v^{\prime \prime }=v^{\prime 3}F(v).
\end{equation*}%
Consequently, the initial third order equation corresponding to Type B is%
\begin{equation}
y^{\prime \prime \prime }=y^{\prime \prime 3}f(y^{\prime }).  \label{5}
\end{equation}

\subsubsection{Type IC}

In this case 
\begin{equation*}
-u^{\prime 3}F(v,\frac{1}{u^{\prime }})=\frac{1}{v}\bar{F}(u^{\prime }),
\end{equation*}%
so that the corresponding third order equation is%
\begin{equation}
y^{\prime \prime \prime }=\frac{1}{y^{\prime }}f(y^{\prime \prime }).
\label{6}
\end{equation}

\subsubsection{Type ID}

The condition 
\begin{equation*}
-u^{\prime 3}F(v,\frac{1}{u^{\prime }})=u^{\prime }\bar{F}(v),
\end{equation*}%
yields%
\begin{equation}
y^{\prime \prime \prime }=y^{\prime \prime 2}f(y)  \label{7}
\end{equation}

\subsection{Type II%
\protect\begin{tabular}{l}
$X=\partial _{y},\text{ \ }Y=x\partial _{y}$%
\protect\end{tabular}%
}

As for Type IA, Type IIA would imply existence of three symmetries in the
initial third order equation, so this case does not belong to the class of
interest. The same applies to Type IIB. Following essentially the same
procedure as for Type I, in a straightforward manner one obtains the
following third order equations:

\subsubsection{Type IIC}

\begin{equation}
y^{\prime \prime \prime }=\frac{1}{x}f(y^{\prime \prime }).  \label{8}
\end{equation}

\subsubsection{Type IID}

\begin{equation}
y^{\prime \prime \prime }=y^{\prime \prime }f(x).  \label{9}
\end{equation}

\subsection{Type III%
\protect\begin{tabular}{l}
$X=\partial _{y},\text{ \ }Y=y\partial _{y}$%
\protect\end{tabular}%
}

The form of (\ref{0}) invariant under $X=\partial _{y}$ is

\begin{equation*}
y^{\prime \prime \prime }=f(x,y^{\prime },y^{\prime \prime }),
\end{equation*}%
while the form of $f$ under the prolongation of $Y$ is obtained applying the
condition%
\begin{equation*}
Y^{(3)}(y^{\prime \prime \prime }-f)\left\vert _{(y^{\prime \prime \prime
}-f)=0}=0\right. ,
\end{equation*}%
yielding%
\begin{equation}
y^{\prime \prime \prime }=y^{\prime \prime }F(x,\frac{y^{\prime \prime }}{%
y^{\prime }}).  \label{10}
\end{equation}%
In terms of the fundamental differential invariants of the group generated
by $X$ and $Y$ eq. (\ref{9}) may be written as%
\begin{equation*}
u^{\prime \prime }=u^{\prime }F(v,u^{\prime })-u^{\prime 2}.
\end{equation*}%
Comparison with the corresponding equations in Table 2. implies that Type
IIIA would require three symmetries in the original third order differential
equation, so this type does not belong to the class of interest here.

\subsubsection{Type IIIB}

From 
\begin{equation*}
u^{\prime }F(v,u^{\prime })-u^{\prime 2}=\bar{F}(v),
\end{equation*}%
since%
\begin{equation*}
ln\text{ }y\prime =u,\ \text{\ \ and}\ \ \ u\prime =\frac{y^{\prime \prime }%
}{y^{\prime }},
\end{equation*}%
it follows in a straightforward manner that the equation we are looking for
is%
\begin{equation}
y^{\prime \prime \prime }=y^{\prime }\left( f(x)+\left( \frac{y^{\prime
\prime }}{y^{\prime }}\right) ^{2}\right) .  \label{11}
\end{equation}

\subsubsection{Type IIIC}

From 
\begin{equation*}
u^{\prime }F(v,u^{\prime })-u^{\prime 2}=\frac{1}{v}\bar{F}(u^{\prime }),
\end{equation*}%
and eq. (\ref{9}) in a straightforward manner one obtains%
\begin{equation}
y^{\prime \prime \prime }=y^{\prime }\left( \frac{1}{x}F\left( \frac{%
y^{\prime \prime }}{y^{\prime }}\right) +\left( \frac{y^{\prime \prime }}{%
y^{\prime }}\right) ^{2}\right) .  \label{12}
\end{equation}

\subsubsection{Type IIID}

From

\begin{equation*}
u^{\prime }F(v,u^{\prime })-u^{\prime 2}=u^{\prime }\bar{F}(v),
\end{equation*}%
the following equation is obtained

\begin{equation}
y^{\prime \prime \prime }=y^{\prime }\left( f(x)\frac{y^{\prime \prime }}{%
y^{\prime }}+\left( \frac{y^{\prime \prime }}{y^{\prime }}\right)
^{2}\right) .  \label{13}
\end{equation}

\subsection{Type IV%
\protect\begin{tabular}{l}
$X=\partial _{y},\text{ \ }Y=x\partial _{x}+y\partial _{y}$%
\protect\end{tabular}%
}

In a manner similar to the one applied to Type III, the form of eq. (\ref{0}%
) invariant under the canonical representation of Type IV algebra is%
\begin{equation}
y^{\prime \prime }=\frac{1}{x^{2}}f(y^{\prime },xy^{\prime \prime }).
\label{14}
\end{equation}%
In terms of the fundamental differential invariants of the group generated
by $X$ and $Y$ eq. (\ref{12}) becomes%
\begin{equation*}
u^{\prime \prime }=-u^{\prime 3}F\left( v,\frac{1}{u^{\prime }}\right)
-u^{\prime 2}.
\end{equation*}%
As in all previous cases Type IA would imply the existence of third symmetry
in the initial third order equation so this case does not belong to the
class in focus of this work.

\subsubsection{Type IVB}

Equation%
\begin{equation*}
u^{\prime \prime }=-u^{\prime 3}F\left( v,\frac{1}{u^{\prime }}\right)
-u^{\prime 2}=\bar{F}(v),
\end{equation*}%
leads to%
\begin{equation*}
F\left( v,\frac{1}{u^{\prime }}\right) =\frac{1}{u^{\prime }}+\frac{1}{%
u^{\prime 3}}\bar{F}(v),
\end{equation*}%
and after straightforward calculation to the equation%
\begin{equation}
y^{\prime \prime \prime }=\frac{y^{\prime \prime }}{x}+xy^{\prime \prime
3}f(y^{\prime }).  \label{15}
\end{equation}

\subsubsection{Type IVC}

From%
\begin{equation*}
u^{\prime \prime }=-u^{\prime 3}F\left( v,\frac{1}{u^{\prime }}\right)
-u^{\prime 2}=\frac{1}{v}\bar{F}(u^{\prime }),
\end{equation*}%
and%
\begin{equation*}
F\left( v,\frac{1}{u^{\prime }}\right) =-\frac{1}{u^{\prime 3}}\left( \frac{1%
}{v}\bar{F}(u^{\prime })+u^{\prime 2}\right) ,
\end{equation*}%
the following equation is obtained%
\begin{equation}
y^{\prime \prime \prime }=-\frac{y^{\prime \prime }}{x}-\frac{xy^{\prime
\prime 3}}{y^{\prime }}F\left( \frac{1}{xy^{\prime \prime }}\right) .
\label{16}
\end{equation}

\subsubsection{Type IVD}

The equation in Table 2. corresponding to Type IV yields 
\begin{equation*}
u^{\prime \prime }=-u^{\prime 3}F\left( v,\frac{1}{u^{\prime }}\right)
-u^{\prime 2}=u^{\prime }\bar{F}(v),
\end{equation*}%
so that the equation of interest is%
\begin{equation}
y^{\prime \prime \prime }=-\frac{y^{\prime \prime }}{x}\left( 1-y^{\prime
\prime }F(y^{\prime })\right) .  \label{17}
\end{equation}

Therefore, the class of third order equations that reduce to second order
ODEs which posses at least one more new point symmetry (possibly a type II
hidden symmetry) includes the equations whose general form is given by
expressions (\ref{5}), (\ref{6}) and (\ref{7}) (Type I), (\ref{8}) and (\ref%
{9}) (Type II), (\ref{10}), (\ref{11}) and (\ref{12}) (Type III) and (\ref%
{15}), (\ref{16}) and (\ref{17}) (Type IV).

\section{Case 2: Reduction to a first order equation possessing one known
(inherited) point symmetry}

This approach uses properties of convertible (hidden) symmetries that arise
along reduction paths of the fourth order differential equations with four
parameter symmetry group that lead to the first order equation possessing
one known symmetry, hence reducible to quadratures. In our approach the
method includes classification of all reduction paths of a fourth order ODE
possessing certain four dimensional algebra of local symmetries to a one
dimensional ODE, and choosing ones that have third order ODE possessing two
parameter symmetry group along one of the paths. Again we stress the fact
that complete classification is not attempted here since the source of
hidden symmetries may be in the equations of order higher than four as well
as in contact symmetries. In the following exposition we use the notation of
reference\cite{Mubarak}, so that $A_{n,m}$ denotes a Lie algebra of
dimension $n$ and isomorphism type $m$. We illustrate the method using
algebras $A_{4,1}$ and $A_{4,4}$.

\subsection{Example 1: Algebra $A_{4,1}$}

The algebra $A_{4,1}$ has a basis%
\begin{equation}
X=\partial _{y},\text{ \ \ \ \ }Y=x\partial _{y},\text{ \ \ \ \ }%
Z=x^{2}\partial _{y},\text{ \ \ \ \ }U=\partial _{x},  \label{18}
\end{equation}%
with nonzero commutator relations%
\begin{equation}
\lbrack Y,U]=X,\text{ \ \ \ \ }[Z,U]=Y.  \label{19}
\end{equation}%
In reference \cite{Nucci} the vector field $U$ is expressed as%
\begin{equation}
U=\partial _{x}+f(x)\partial _{y}.  \label{20}
\end{equation}%
However it may be noticed that since canonical coordinates are not uniquely
defined and satisfy transformations%
\begin{eqnarray}
\bar{x} &=&F(x)  \label{21} \\
\bar{y} &=&y+G(x),  \notag
\end{eqnarray}%
for arbitrary smooth functions $F$ and $G$ (with additional constraint $%
F^{^{\prime }}(x)\neq 0),$ expression (\ref{20}) may be reduced to the form
in (\ref{18}). Consequently, the most general invariant fourth order
equation admitting the Lie symmetry algebra A$_{4,1}$ has a simpler form%
\begin{equation}
\bigtriangleup ^{4}:y^{iv}=F(y^{\prime \prime \prime }),  \label{22}
\end{equation}%
where $\bigtriangleup ^{n}$ denotes a differential equation of order $n$.
Although eq. (\ref{22}) may be easily integrated without the use of symmetry
methods, it is instructive to analyze the reduction paths for this equation.
The reduction scheme of A$_{4,1}$ algebra is presented in Table A.1 of
Appendix A. In the notation used superscript $N$ denotes nonlocal symmetry
and tilde denotes restriction of the inherited symmetry generator to
corresponding fundamental differential invariants, while the superscript in
the parenthesis denotes the order of the prolongation.

In the construction of the reduction paths$,$ well known conditions for
disappearance, preservation and reappearance of point symmetries have been
used. The particular reduction path chosen for clarification of the method
is presented below in a schematic manner.

\bigskip

\begin{equation*}
\begin{tabular}{ccc}
& $\bigtriangleup ^{4}$: A$_{4,1}(X,Y,Z,U)$ &  \\ 
& $\Downarrow Z$ &  \\ 
$\swarrow $ & $\bigtriangleup ^{3}$: A$_{2,1}(\tilde{X}^{(1)};\tilde{Y}%
^{(1)});\tilde{U}^{(1),N}$ & $\searrow $ \\ 
$\Downarrow \tilde{X}^{(1)}$ &  & $\tilde{Y}^{(1)}\Downarrow $ \\ 
$\bigtriangleup ^{2}$: \ A$_{1}(\tilde{Y}^{(2)});\tilde{U}^{(2),N}$ &  & $%
\bigtriangleup ^{2}$: \ A$_{1}(\tilde{X}^{(2)});\tilde{U}^{(2),N}$ \\ 
$\Downarrow \tilde{Y}^{(2)}$ &  & $\Downarrow \tilde{X}^{(2)}$ \\ 
$\bigtriangleup ^{1}$: A$_{1}(\tilde{U}^{(3)})$ &  & $\bigtriangleup ^{1}$:
\ A$_{1}(\tilde{U}^{(3)})$%
\end{tabular}%
\end{equation*}

\bigskip

In terms of the fundamental differential invariants $r$ and $v$ of the group
generated by $Z$, equation (\ref{22}) becomes%
\begin{equation}
r^{2}v^{^{\prime \prime \prime }}+8rv^{\prime \prime }+12v^{\prime
}=F(r^{2}v^{\prime \prime }+6rv^{\prime }+6v).  \label{23}
\end{equation}%
Restrictions of the remaining vector fields in terms of $r$ and $v$ are:%
\begin{eqnarray*}
\tilde{X}^{(1)} &=&-\frac{2}{r^{3}}\partial _{v}, \\
\tilde{Y}^{(1)} &=&-\frac{1}{r^{2}}\partial _{v}, \\
\tilde{U}^{(1),N} &=&\partial _{r}+\left( r^{2}\dint vdr-\frac{2v}{r}\right)
\partial _{v}.
\end{eqnarray*}%
In terms of the fundamental differential invariants of symmetry $\tilde{X}%
^{(1)}$ which we denote as $x$ and $y$, 
\begin{eqnarray*}
x &=&r, \\
y &=&-\frac{1}{2}r^{3}v.
\end{eqnarray*}%
A straightforward calculation turns equation (\ref{22}) into 
\begin{equation}
y^{\prime \prime \prime }=\frac{y^{\prime \prime }}{x}+xF\left( \frac{%
y^{\prime \prime }}{x}\right) .  \label{24}
\end{equation}%
This equation has two point symmetries $X=\partial _{y},$\ $Y=x\partial _{y}$
and reduces to quadrature as shown in the diagram above. Naturally, as
presented in the schematic form above, reducing the order of a third order
equation in the reduction path using symmetry $\tilde{Y}^{(1)}$ first also
leads to a first order equation solvable by quadrature. In terms of the
fundamental differential invariants of symmetry $\tilde{Y}^{(1)}=-\frac{1}{%
r^{2}}\partial _{v},$ denoted by $\rho $ and $\theta ,$ an equation also
possessing two parameter symmetry group is obtained however it is of
slightly more complicated form. A straightforward calculation for this case
yields%
\begin{eqnarray*}
\rho &=&r, \\
\theta &=&-r^{2}v,
\end{eqnarray*}%
so that with $w=\theta ^{\prime }$ equation (\ref{22}) becomes%
\begin{equation}
w^{\prime \prime }+2\frac{w^{\prime }}{\rho }-2\frac{w}{\rho ^{2}}=F\left(
w^{\prime }+2\frac{w}{\rho }\right) .  \label{25}
\end{equation}%
The restriction of the inherited symmetry to the fundamental differential
invariants is%
\begin{equation*}
\tilde{X}^{(2)}=-\frac{2}{\rho ^{2}}\partial _{w},
\end{equation*}%
while the nonlocal symmetry is%
\begin{equation*}
\tilde{U}^{(2),N}=\partial _{\rho }+\frac{2}{\rho ^{2}}\dint wd\rho \text{ }%
\partial _{w}.
\end{equation*}%
Further reduction using $\tilde{X}^{(2)}$ yields the following first order
equation%
\begin{equation}
y^{\prime }-2\frac{y}{x}=x^{2}F\left( \frac{y}{x^{2}}\right) ,  \label{26}
\end{equation}%
whose symmetry is%
\begin{equation*}
\tilde{U}^{(3)}=\partial _{x}+2\frac{y}{x}\partial _{y}.
\end{equation*}%
Equation (\ref{26}) is of Riccati type and easily solvable by quadrature.

\subsection{Example 2: Algebra $A_{4,4}$}

The algebra $A_{4,4}$ has a basis%
\begin{equation}
X=\partial _{y},\text{ \ \ \ \ }Y=x\partial _{y},\text{ \ \ \ \ }%
Z=x^{2}\partial _{y},\text{ \ \ \ \ }U=-\partial _{x}+y\partial _{y},
\label{27}
\end{equation}%
with nonzero commutator relations%
\begin{equation}
\lbrack X,U]=X,\text{ \ \ \ \ }[Y,U]=X+U,\text{ \ \ \ \ }[Z,U]=Y+Z.
\label{28}
\end{equation}%
The most general invariant fourth order equation admitting the Lie symmetry
algebra A$_{4,4}$ is 
\begin{equation}
\bigtriangleup ^{4}:y^{iv}=e^{-x}\text{ }F(e^{x}y^{\prime \prime \prime }).
\label{29}
\end{equation}%
From the reduction paths presented in the Appendix, the only reduction path
of interest is presented below in a schematic manner.

\bigskip

\begin{equation*}
\begin{tabular}{ccc}
& $\bigtriangleup ^{4}$: A$_{4,4}(X,Y,Z,U)$ &  \\ 
& $\Downarrow Z$ &  \\ 
$\swarrow $ & $\bigtriangleup ^{3}$: A$_{2,1}(\tilde{X}^{(1)};\tilde{Y}%
^{(1)});\tilde{U}^{(1),N}$ & $\searrow $ \\ 
$\Downarrow \tilde{X}^{(1)}$ &  & $\tilde{Y}^{(1)}\Downarrow $ \\ 
$\bigtriangleup ^{2}$: \ A$_{1}(\tilde{Y}^{(2)});\tilde{U}^{(2),N}$ &  & $%
\bigtriangleup ^{2}$: \ A$_{1}(\tilde{X}^{(2)});\tilde{U}^{(2),N}$ \\ 
$\Downarrow \tilde{Y}^{(2)}$ &  & $\Downarrow \tilde{X}^{(2)}$ \\ 
$\bigtriangleup ^{1}$: A$_{1}(\tilde{U}^{(3)})$ &  & $\bigtriangleup ^{1}$:
\ A$_{1}(\tilde{U}^{(3)})$%
\end{tabular}%
\end{equation*}

\bigskip

Calculations for this algebra are very similar to the ones for algebra A$%
_{4,1}$ so that the third order equation possessing two parameter symmetry
group reducible to quadratures is 
\begin{equation*}
y^{\prime \prime \prime }=\frac{y^{\prime \prime }}{x}+xe^{-x}F\left( e^{x}%
\frac{y^{\prime \prime }}{x}\right) .
\end{equation*}%
Following a straightforward calculation the first order equation
corresponding to the eq. (\ref{26}) is%
\begin{equation*}
y^{\prime }-2\frac{y}{x}=x^{2}e^{-x}F(e^{x}\frac{y}{x^{2}}).
\end{equation*}

{\LARGE Appendix \bigskip A}

\appendix{}

Reduction paths for four parameter symmetry groups are presented in Tables
1-21. In the notation used superscript $N$ denotes nonlocal symmetry and
tilde denotes restriction of the inherited symmetry generator to the
corresponding fundamental differential invariants, while the superscript in
the parenthesis denotes the order of the prolongation. Known conditions for
disappearance, preservation and reappearance of point symmetries have been
used in the construction of reduction paths along with conditions derived
and discussed in \cite{Nikolic}. The reduction paths along which third order
equations having two parameter symmetry group may be reduced to quadratures
are marked by black triangles.

\end{document}